\documentclass[a4paper]{article}
\usepackage[colorlinks]{hyperref}

\usepackage{amsmath,latexsym,amssymb}
\usepackage{array}
\usepackage[all]{xy}

\newcolumntype{C}[1]{>{\centering\arraybackslash}p{#1}}

\newcommand*{\myforall}[1]{\mathop{\forall#1.\,}\nolimits}
\newcommand*{\myforsome}[1]{\mathop{\exists#1.\,}\nolimits}
\newcommand*{\dom}[2]{\mathrel{:}{{#1}\rightarrow{#2}}}
\newcommand*{\blank}{\mathord{-}}

\newcommand*{\comp}{\mathbin{\circ}}

\newcommand*{\nattr}{\mathrel{\trianglelefteq}}
\newcommand*{\revnattr}{\mathrel{\trianglerighteq}}
\newcommand*{\inv}[1]{#1^{-1}}
\newcommand*{\opp}[1]{#1^{\mathrm{op}}}
\newcommand*{\cset}[1]{\lbrace#1\rbrace}
\newcommand*{\sset}[2]{\cset{\,{#1}\mathrel{\mid}{#2}\,}}
\newcommand*{\tuple}[1]{\langle#1\rangle}
\newcommand*{\union}{\mathop{\mathchoice{\textstyle\bigcup}{\bigcup}{\bigcup}{\bigcup}}}

\newcommand*{\join}{\mathop{\mathchoice{{\textstyle\bigvee}}{\bigvee}{\bigvee}{\bigvee}}}
\newcommand*{\meet}{\mathop{\mathchoice{{\textstyle\bigwedge}}{\bigwedge}{\bigwedge}{\bigwedge}}}
\newcommand*{\copower}{\mathbin{\cdot}}
\newcommand*{\exponent}[2]{#2^{#1}}
\newcommand*{\modal}[1]{[{#1}]}
\newcommand*{\dmodal}[1]{\langle{#1}\rangle}

\newcommand*{\curry}[1]{{\widetilde{#1}}}
\newcommand*{\lift}[1]{#1^{\ast}}
\newcommand*{\lowera}[1]{{\downarrow}_{#1}}
\newcommand*{\uppera}[1]{{\uparrow}_{#1}}
\newcommand*{\eleq}[1]{\leq_{#1}}
\newcommand*{\interval}[2]{[{#1},{#2}]}

\newcommand*{\sets}{\textrm{\bf Set}}
\newcommand*{\psets}{\textrm{\bf PSet}}

\newcommand*{\cats}[1]{{#1}\textrm{\bf -Cat}}

\newcommand*{\mathconj}[1]{\mathrel{\;#1\;}}
\newcommand*{\If}{\mathconj{\Leftarrow}}
\newcommand*{\Oif}{\mathconj{\Rightarrow}}
\newcommand*{\Iff}{\mathconj{\Leftrightarrow}}
\newcommand*{\textif}{\mathconj{\text{if}}}

\newcommand*{\otherwise}{\mathconj{\text{otherwise}}{}}
\newcommand*{\where}{\mathconj{\text{where}}}

\usepackage{amsthm}
\theoremstyle{plain}
\providecommand{\newdefinition}[2]{\theoremstyle{definition}\newtheorem{#1}{#2}\theoremstyle{plain}}
\newtheorem{theorem}{Theorem}

\newtheorem{lemma}[theorem]{Lemma}

\newdefinition{definition}{Definition}
\newdefinition{example}{Example}
\newdefinition{remark}{Remark}

\begin{document}

\title{A Categorical Foundation of Rough Sets}

\author{Yoshihiko Kakutani\thanks{
  No affiliation (Most of this work had been studied until 2019 at University of Tokyo.) \\
  e-mail: \texttt{yoshihiko.kakutani@gmail.com}
 }}
\date{2025-04-05}

\maketitle

\begin{abstract}
 Rough sets are approximations of concrete sets.
 The theory of rough sets has been used widely for data-mining.
 While it is well-known that adjunctions are underlying in rough approximations, such adjunctions are not enough for characterization of rough sets.
 This paper provides a way to characterize rough sets in terms of category theory.
 We reformulate rough sets as adjunctions between preordered sets in a general way.
 Our formulation of rough sets can enjoy benefits of adjunctions and category theory.
 Especially, our characterization is closed under composition.
 We can also explain the notions of attribute reduction and data insertion in our theory.
 It is novel that our theory enables us to guess decision rules for unknown data.
 If we change the answer set, we can get a refinement of rough sets without any difficulty.
 Our refined rough sets lead rough fuzzy sets or more general approximations of functions.
 Moreover, our theory of rough sets can be generalized in the manner of enriched category theory.
 The derived enriched theory covers the usual theory of fuzzy rough sets.
\end{abstract}

\section{Introduction}

The concept of rough sets was proposed by Pawlak~\cite{Pawlak82:RS} in order to deal with vagueness.
While a crisp set has a clear boundary, a rough set has a vague boundary.
Mathematically, a rough set consists of two kinds of approximations, an upper approximation and a lower one.
The region between upper and lower approximations represents the boundary of a property.
Rough sets have been studied constantly in the field of data-mining, and recently, much attention is paid to their practical applications.
Both in a theoretical aspect and in a practical aspect, we can refer to kinds of extensions of rough sets~\cite{DuboisPrade90:RFSFRS,LinLiu94:RAO,Yao98:CAMTRS,Kryszkiewicz98:RSAIIS,PawlakSkowron07:RS,QinYang08:GRSBRTR,KandilTantawy14:GRSTOS,ZhangLi14:CRSDDM}.

Although rough approximations can be calculated directly, we formulate them with adjunctions in this work.
An adjunction between preordered sets is often referred to as a Galois connection.
Some Galois connections in rough set theory~\cite{Jarvinen02:SRA,DuentschGediga03:AOQDA} were studied but are not enough for categorical reasoning of rough set theory.
Our adjunctions are division of a known adjunction and explain how approximations are categorically constructed from original data.

In rough set theory, data are given as a collection of objects.
Each object has attributes and is classified into `yes' or `no'.
Our analysis is aimed at characterizing conditions on attributes for the classification.
When objects $x$ and $y$ have the same attributes, we write $x \equiv y$.
Clearly, ${\equiv}$ is an equivalence relation.
Let $\mu$ be the set of all objects classified into yes.
The lower approximation of $\mu$ means the subset of ${X}/{\equiv}$ consisting of classes surely in yes.
On the other hand, the concept of the upper approximation of $\mu$ is the subset possibly in yes.
These approximations are the essential concept of a rough set.

In our formulation, the lower approximation is characterized as the right adjoint of the trivial inclusion, and dually, the upper approximation is characterized as the left adjoint.
In some axiomatizations~\cite{LinLiu94:RAO,Yao98:CAMTRS}, an upper approximation is the dual of a lower approximation via a kind of negation operator.
In our framework, the duality is represented as the symmetry between left and right adjoints.

Given an equivalence relation on objects, there are two adjunctions, right and left adjoints, in our formalization.
From the composite of these two adjunctions, a known Galois connection~\cite{Jarvinen02:SRA} can be derived.
Our theory has some advantages over other category theoretic approaches.
First, approximations are closed under composition in our theory.
In other words, we have a lower/upper approximation correctly even if we approximate more than once.
Second, the notions of attribute reduction and data insertion can be explained in a sophisticated way.
Third, we can guess decision rules for unknown data automatically.
While the usual rough set theory can be derived with a surjective functor in our theory, a non-surjective functor enables us to guess decision rules.
We show some examples for the above points in the paper.

Since our framework follows quite standard approach in category theory, we can generalize the theory easily.
While a set of objects is usually discrete, we can consider a preorder on it.
Though such a generalization may not be required by practical applications, it makes the theory clear and coherent.
For another extension, we can replace the answer set of `yes' and `no' with a more complex structure.
Moreover, we can apply enriched category theory to our framework.
This enriched version covers the usual theory of fuzzy rough sets~\cite{DuboisPrade90:RFSFRS}.

Fuzzy sets~\cite{Zadeh65:FS} have been studied widely as another approach to vagueness.
A fuzzy set has an extension of a membership function with intermediate values.
From the early days of the history of rough sets, similarities and differences between fuzzy sets and rough sets were discussed~\cite{Pawlak82:RSFS,Yao98:CSFSRS}.
Fusion of rough sets and fuzzy sets is known as fuzzy rough sets~\cite{DuboisPrade90:RFSFRS}.
Theories of fuzzy rough sets are usually provided algebraically.
This paper gives another reasoning to fuzzy rough sets via enriched category theory.

From the viewpoint of enriched category theory, a preordered set is a $2$-enriched category and a monotone function between preordered sets is a $2$-enriched functor, where $2$ is the trivial lattice $\cset{\bot,\top}$.
Therefore, we replace $2$ with a complete commutative residuated lattice $L$.
Significance of residuated lattices in generalizations of fuzzy rough sets was already known~\cite{RadzikowskaKerre04:FRSBRL,SheWang09:AAFRSBRL}.
If we take the interval between $0$ and $1$ as a residuated lattice, we can get the usual theory of fuzzy rough sets from our theory.
More general structures like quantaloids~\cite{Stubbe05:CSEQ,Stubbe06:CSEQ} can be considered for enrichment.
We do not focus on such generalization, but we expect that most part of this work can be generalized.

\subsection*{Related Work}

While we are formalizing the theory in terms of category theory, most studies on rough sets are directed to characterize the theory as a set of algebraic axioms.
Although the studies on Galois connections~\cite{Jarvinen02:SRA,DuentschGediga03:AOQDA} are strongly related to our work, their aim is still an algebraic characterization.
Relation of Galois connections and rough sets is studied also in formal concept analysis~\cite{Wille82:RLT,DuentschGediga03:AOQDA,Wolski04:GCDA,Yao04:CSFCARSTDA}.
Such studies provides another direction of generalization of rough set theory.
The intersection of formal concept analysis and our work is just the traditional rough set theory.

The approach to generalize rough sets to fuzzy rough sets~\cite{RadzikowskaKerre02:CSFRS,WuMi03:GFRS,RadzikowskaKerre04:FRSBRL,SheWang09:AAFRSBRL,MaHu13:TLSFRSDLUS} is similar to our enrichment method.
In those studies, a relation is replaced with a fuzzy relation over a residuated lattice.
Some aspects of such generalization can be explained in terms of enriched category theory.
Also fuzzy extensions~\cite{Belohlavek04:CLOFL,ShaoLiu07:SAFFCA,LaiZhang09:CLFC} of formal concept analysis are studied algebraically.
Studies on fuzzy Galois connections~\cite{Belohlavek99:FGC,YaoLu09:FGCFP} have more direct relation to our enrichment of adjunctions.
A complete residuated lattice is sometimes referred to as a unital quantale, which is an instance of quantaloids.
Enrichment by a quantaloid can be useful in various fuzzy theories.
Especially, Shen's work~\cite{Shen14:AQEC} provides quantaloid-enriched adjunctions along the line of formal concept analysis.
For the enrichment itself, it goes without saying that enrichment by quantaloids is more general than ours mathematically.

There are some other important approaches to fuzzy rough set theory.
Above all, topological properties of approximations~\cite{QinPei05:TPFRS,LaiZhang06:FPFT,HaoLi11:RLFRSLT,YuZhan14:TPGRS,Wang17:TCGFRS} are studied actively.
The relation between topological approach and algebraic approach is studied well, but it is not clear how topological properties are related to our study.

\subsection*{Construction}

This paper is constructed as follows.

In Section~\ref{Sec:Simple}, we give the definition of rough sets in terms of category theory.
In our framework, approximations are defined as adjunctions between preordered sets.
In Section~\ref{Sec:Concepts}, it is demonstrated with examples that our theory is a generalization of the usual theory of rough sets.
The notions of attribute reduction and data insertion are characterized in our framework.
In addition, inference of decision rules for unknown data is discussed.
Next, we provide a refinement of our theory of rough sets in Section~\ref{Sec:Refined}.
In this refined theory, not only approximations of sets but also approximations of functions can be considered.
Last, the enriched theory of rough sets is discussed in Section~\ref{Sec:Enriched}.
Our enriched theory is a generalization of the theory given in Section~\ref{Sec:Simple}, based on enriched category theory.
Our enriched theory gives categorical reasoning of fuzzy rough sets theory.

It should be noted that most proofs of the theorems are omitted in this paper.
It is the reason that there are well-known theorems more general than ours.
One can refer to textbooks for details: e.g., Mac Lane's~\cite{MacLane71:CWM} for Section~\ref{Sec:Simple}, \ref{Sec:Concepts} and \ref{Sec:Refined}, Kelly's~\cite{Kelly82:BCECT} for Section~\ref{Sec:Enriched}.
This paper, however, do not require the knowledge of category theory.
For a technical term, another equivalent statement is provided in the paper.

\section{Abstract Rough Sets}
\label{Sec:Simple}

In this paper, $2$ denotes the lattice $\cset{\bot,\top}$ with $\bot < \top$.
For any preordered set $X$, its power set $2^{X}$ is also a preordered set.
Given a monotone (order-preserving) function $R \dom{X}{A}$, $\lift{R} \dom{2^{A}}{2^{X}}$ is defined by $\lift{R}{\mu} = {\mu}\comp{R}$.
In this section, we regard a preordered set as a category and a monotone function as a functor.
The category of preordered sets is denoted by $\psets$.

\begin{definition}
 For a functor $R \dom{X}{A}$ between preordered sets, the \emph{upper approximation} $\uppera{R} \dom{2^{X}}{2^{A}}$ is a left adjoint of $R^{\ast}$, the \emph{lower approximation} $\lowera{R} \dom{2^{X}}{2^{A}}$ is a right adjoint of $R^{\ast}$.
\end{definition}

Explicit descriptions of the definition are as follows.
An adjunction between preordered sets is often called a \emph{Galois connection}.
\begin{itemize}
 \item For any $\nu \in{2^{A}}$, $\uppera{R}{\mu} \subset \nu$ is equivalent to $\mu \subset \lift{R}{\nu}$.
 \item For any $\nu \in{2^{A}}$, $\nu \subset \lowera{R}{\mu}$ is equivalent to $\lift{R}{\nu} \subset \mu$.
\end{itemize}

In general, a left/right adjoint functor is uniquely determined up to isomorphisms.
Therefore, our approximations are well-defined.
It is well-known that approximations always exist.

\begin{theorem}\label{Thm:Ex}
 The upper/lower approximation exists uniquely for any $R \dom{X}{A}$.
 The pointwise definitions of the approximations are as follows.
 \begin{align*}
  \uppera{R}{\mu} &= \sset{a}{\inv{R}(A_{\leq{a}}) \cap \mu \neq \emptyset}
  & &\where A_{\leq{a}} = \sset{b}{b \leq a} \\
  \lowera{R}{\mu} &= \sset{a}{\inv{R}(A_{\geq{a}}) \subset \mu}
  & &\where A_{\geq{a}} = \sset{b}{b \geq a}
 \end{align*}
\end{theorem}

It may not be clear that the above definition coincides with the usual definition of rough sets.
For a functor $R \dom{X}{A}$, let ${\equiv}$ be the equivalence relation on $X$ such that $x \equiv y$ is equivalent to $R{x} = R{y}$.
Here, we assume that $X$ and $A$ are discrete sets.
Then, the image of $R$ is isomorphic to $X/{\equiv}$.
Especially, $A$ is isomorphic to $X/{\equiv}$ if $R$ is surjective.
In this case, $\lift{R}$ is essentially the same as the inclusion from $2^{{X}/{\equiv}}$ to $2^{X}$, and $\lift{R}(\lowera{R}{\mu}) = (\lowera{R}{\mu})\comp{R} \in{2^{X}}$ is a lower approximation of $\mu$ in the usual sense.
Under the same condition, $\lift{R}(\uppera{R}{\mu})$ is an upper approximation in the usual sense.
Therefore, our formulation covers the usual approximations in rough set theory.

We show an example of approximations to see validity of our formulation.
As seen in the following example, it is a typical case that $A$ is a product (or its subset) of some attribute sets.

\begin{example}\label{Ex:Simple}
 Consider the following test data.
 \begin{center}
  \begin{tabular}[t]{c|c|c|c}
   object & attribute 1 & attribute 2 & answer \\ \hline \hline
   $0$  & red    & low  & yes \\ \hline
   $1$  & red    & low  & yes \\ \hline
   $2$  & blue   & low  & no  \\ \hline
   $3$  & yellow & low  & no  \\ \hline
   $4$  & yellow & low  & yes \\ \hline
   $5$  & white  & low  & no  \\ \hline
   $6$  & red    & high & yes \\ \hline
   $7$  & blue   & high & no  \\ \hline
   $8$  & blue   & high & no  \\ \hline
   $9$  & yellow & high & yes \\ \hline
   $10$ & white  & high & yes \\ \hline
   $11$ & white  & high & no
  \end{tabular}
 \end{center}
 Let $X$ and $A$ be discrete sets $\cset{0,1,\dots,11}$ and $\cset{\text{red},\text{blue},\text{yellow},\text{white}}\times\cset{\text{low},\text{high}}$.
 If we focus on the answer ``yes'', $\mu$ is $\cset{0,1,4,6,9,10} \in{2^{X}}$.
 Since $R$ is also defined from the table, the approximations can be calculated.
 \begin{align*}
  \uppera{R}{\mu} &=
  \cset{(\text{red},\text{low}),(\text{yellow},\text{low}),
   (\text{red},\text{high}),(\text{yellow},\text{high}),(\text{white},\text{high})} \\
  \lowera{R}{\mu} &=
  \cset{(\text{red},\text{low}),(\text{red},\text{high}),(\text{yellow},\text{high})}
 \end{align*}
 If we use the usual representation, the following results are obtained.
 \begin{align*}
  \lift{R}(\uppera{R}{\mu}) &=
  \cset{0,1,3,4,6,9,10,11} \\
  \lift{R}(\lowera{R}{\mu}) &=
  \cset{0,1,6,9}
 \end{align*}
\end{example}

In this paper, the term `upper approximation' refers to not only $\uppera{R}$ but also $\uppera{R}{\mu}$ and $\lift{R}(\uppera{R}{\mu})$, following the usual terminology of rough set theory.
We use the same terminology for lower approximations.

\begin{remark}
 Since a composite of adjunctions is also an adjunction, ${\lift{R}}\comp{\uppera{R}} \dom{2^{X}}{2^{X}}$ is a left adjoint of ${\lift{R}}\comp{\lowera{R}} \dom{2^{X}}{2^{X}}$.
 This adjunction is regarded as an important property of approximations in rough set theory~\cite{Jarvinen02:SRA,DuentschGediga03:AOQDA}.
 While information of $R$ are lost in this composite, our theory provides categorical construction of the lower/upper approximation from $R$.
\end{remark}

\begin{remark}
 According to the general theory of adjunctions, ${\lift{R}}\comp{\uppera{R}}$ is a monad and ${\lift{R}}\comp{\lowera{R}}$ is a comonad on $2^{X}$.
 A monad/comonad on a preordered set is always idempotent.
 Let $\Diamond$ be ${\lift{R}}\comp{\uppera{R}}$ and $\Box$ be ${\lift{R}}\comp{\lowera{R}}$.
 A $\Diamond$-algebra is an object $\mu$ such that $\Diamond{\mu} = \mu$ holds.
 (Or, equivalently, a $\Diamond$-algebra has the form $\Diamond{\mu}$.)
 The category of $\Diamond$-algebras, which is called the Eilenberg-Moore category of $\Diamond$, is isomorphic to $2^{{X}/{\equiv}}$, where $\equiv$ is mentioned above.
 Dually, the Eilenberg-Moore category of $\Box$ is also $2^{{X}/{\equiv}}$.
\end{remark}

\begin{remark}
 In studies on formal concept analysis~\cite{DuentschGediga03:AOQDA,Wolski04:GCDA,Yao04:CSFCARSTDA}, $\lift{R}$ may be referred to as $\modal{\inv{R}}$ and $\dmodal{\inv{R}}$.
 We do not take such notations because the direction of our generalization is different from formal concept analysis.
\end{remark}

\section{Properties of Rough Sets}
\label{Sec:Concepts}

In this section, we show some notions on rough sets can be expressed in our framework.
Properties of rough sets can be derived from general theorems in category theory.

We know that a right/left adjoint preserves limits/colimits.
Because an intersection is a product and a union is a coproduct in $2^{X}$, the following properties follow immediately.

\begin{theorem}\label{Thm:Prsrv}
 For any objects $\mu$ and $\nu$ in $2^{X}$, $\lowera{R}({\mu}\cap{\nu}) = (\lowera{R}{\mu})\cap(\lowera{R}{\nu})$ and $\uppera{R}({\mu}\cup{\nu}) = (\uppera{R}{\mu})\cup(\uppera{R}{\nu})$ hold.
\end{theorem}

One of the advantages of our formulation is compositionality of approximations.
The following theorem guarantees that approximations are closed under composition.
The proof is trivial because of the uniqueness and compositionality of adjunctions.

\begin{theorem}\label{Thm:Comp}
 For any functors $R \dom{X}{A}$ and $R' \dom{A}{A'}$, $\uppera{R'}(\uppera{R}{\mu}) = \uppera{{R'}\comp{R}}{\mu}$ and $\lowera{R'}(\lowera{R}{\mu}) = \lowera{{R'}\comp{R}}{\mu}$ hold.
\end{theorem}

We can define reducibility of attributes as follows.
The definition is just saying that $R$ and ${R'}\comp{R}$ give the same approximations of $\mu$.
We can remark that $\uppera{R}{\mu} \subset \lift{R'}(\uppera{{R'}\comp{R}}{\mu})$ and $\lift{R'}(\lowera{{R'}\comp{R}}{\mu}) \subset \lowera{R}{\mu}$ are always true.

\begin{definition}
 Given $R \dom{X}{A}$ and $\mu \in{2^{X}}$.
 $A$ is \emph{reducible} to $A'$ along $R' \dom{A}{A'}$ if $\lift{R'}(\uppera{{R'}\comp{R}}{\mu}) = \uppera{R}{\mu}$ and $\lift{R'}(\lowera{{R'}\comp{R}}{\mu}) = \lowera{R}{\mu}$ hold.
\end{definition}

When $R'$ is an inclusion in the definition, $A'$ can be considered an extension of $A$ rather than a reduced space.
So, in this case, we may say that $R'$ is \emph{conservative}.
In the usual setting, $R'$ is a projection between subsets of product spaces like the following example.

\begin{example}
 Given the following table, which is an extension of the table in Example~\ref{Ex:Simple}.
 \begin{center}
  \begin{tabular}[t]{c|c|c|c|c}
   object & attribute 1 & attribute 2 & attribute 3 & answer \\ \hline \hline
   $0$  & red    & low  & iris    & yes \\ \hline
   $1$  & red    & low  & freesia & yes \\ \hline
   $2$  & blue   & low  & daisy   & no  \\ \hline
   $3$  & yellow & low  & daisy   & no  \\ \hline
   $4$  & yellow & low  & daisy   & yes \\ \hline
   $5$  & white  & low  & iris    & no  \\ \hline
   $6$  & red    & high & freesia & yes \\ \hline
   $7$  & blue   & high & daisy   & no  \\ \hline
   $8$  & blue   & high & iris    & no  \\ \hline
   $9$  & yellow & high & freesia & yes \\ \hline
   $10$ & white  & high & daisy   & yes \\ \hline
   $11$ & white  & high & daisy   & no
  \end{tabular}
 \end{center}
 In this case, $X$ and $\mu$ are the same as Example~\ref{Ex:Simple}.
 Let $A'$ and $R'$ refer to the table, where $A'$ consists of $10$ elements of $\cset{\text{red},\text{blue},\text{yellow},\text{white}}\times\cset{\text{low},\text{high}}\times\cset{\text{iris},\text{freesia},\text{daisy}}$.
 The upper and lower approximations w.r.t.\ $R'$ coincide with the approximations of Example~\ref{Ex:Simple}.
 \begin{align*}
  \lift{R'}(\uppera{R'}{\mu}) &=
  \cset{0,1,3,4,6,9,10,11} \\
  \lift{R'}(\lowera{R'}{\mu}) &=
  \cset{0,1,6,9}
 \end{align*}
 We can define $p$ as the restriction of the projection from $\cset{\text{red},\text{blue},\text{yellow},\text{white}}\times\cset{\text{low},\text{high}}\times\cset{\text{iris},\text{freesia},\text{daisy}}$ to $\cset{\text{red},\text{blue},\text{yellow},\text{white}}\times\cset{\text{low},\text{high}}$.
 Since ${p}\comp{R'}$ is just $R$ in Example~\ref{Ex:Simple} and $R'$ is epic, $A'$ is reducible to $A$ along $p$.
\end{example}

Reducibility can be characterized algebraically as follows.
This characterization is sometimes used in practical studies.

\begin{theorem}\label{Thm:RedChar}
 Given $R \dom{X}{A}$ and $\mu \in{2^{X}}$.
 $A$ is reducible to $A'$ along $R'$ if and only if $R'{a} \leq R'{b}$ implies $(\uppera{R}{\mu}){a} \leq (\uppera{R}{\mu}){b}$ and $(\lowera{R}{\mu}){a} \leq (\lowera{R}{\mu}){b}$.
\end{theorem}

By Theorem~\ref{Thm:Comp}, it can be seen easily that the reducibility relation is transitive.
Since $A$ is reducible to $A$ itself, reducibility functors compose a subcategory of $\psets$.
In data-mining, it is important to identify a preordered set no more reducible along a projection.

\begin{theorem}\label{Thm:RedComp}
 Given $R \dom{X}{A}$ and $\mu \in{2^{X}}$.
 If $A$ is reducible to $A'$ along $R'$ and $A'$ is reducible to $A''$ along $R''$, $A$ is reducible to $A''$ along ${R''}\comp{R'}$.
\end{theorem}

We have seen the case to shift the codomain of $R$.
Next, we change the domain of $R$ in order to represent data update.
Our notion of update supports not overwriting original values but inserting new data.

\begin{definition}
 A morphism in the comma category ${\psets}/({A}\times{2})$ is called an \emph{update}.
 A functor $i \dom{X}{X'}$ is an update from $\tuple{R,\mu}$ to $\tuple{R',\mu'}$ if and only if ${R'}\comp{i} = R$ and ${\mu'}\comp{i} = \mu$ hold.
\end{definition}

Typically, an update morphism is an inclusion.
Also merging the same data is an update in our sense.
Update morphisms are closed under composition by definition.

\begin{example}
 Let $X$, $A$, $R$ and $\mu$ be the same as in Example~\ref{Ex:Simple}.
 Consider the following additional data with Example~\ref{Ex:Simple}.
 \begin{center}
  \begin{tabular}[t]{c|c|c|c}
   object & attribute 1 & attribute 2 & answer \\ \hline \hline
   $12$ & blue & high & yes \\ \hline
   $13$ & yellow & high & no  \\ \hline
   $14$ & white & high & yes
  \end{tabular}
 \end{center}
 Define $X'$, $R'$, and $\mu'$ from the data.
 Then, $X'$ is ${X}\cup\cset{12,13,14}$, and the usual inclusion $i \dom{X}{X'}$ is an update from $\tuple{R,\mu}$ to $\tuple{R',\mu'}$.
 The approximations of $\mu'$ can be calculated as follows.
 \begin{align*}
  \uppera{R'}{\mu'} &=
  {\uppera{R}{\mu}}\cup\cset{(\text{blue},\text{high})} \\
  \lowera{R'}{\mu'} &=
  {\uppera{R}{\mu}}\setminus\cset{(\text{yellow},\text{high})}
 \end{align*}
 Let $X''$ be ${X'}\setminus{\cset{1,8}}$.
 $R''$ and $\mu''$ can be defined canonically as restrictions of $R'$ and $\mu'$.
 We can define $i' \dom{X'}{X''}$ so that $i'{1} = 0$, $i'{8} = 7$ and $i'{x} = x$ for other $x$ hold.
 This $i'$ is an update from $\tuple{R',\mu'}$ to $\tuple{R'',\mu''}$.
 In this case, clearly, $\uppera{R''}{\mu''} = \uppera{R'}{\mu'}$ and $\lowera{R''}{\mu''} = \lowera{R'}{\mu'}$ hold.
\end{example}

Boundary between lower and upper approximations may become wider after an update.
We can see this fact mathematically.
Let $i \dom{X}{X'}$ be an update from $\tuple{R,\mu}$ to $\tuple{R',\mu'}$.
$\uppera{R}{\mu} \subset \uppera{R'}{\mu'}$ can be derived as follows.
\begin{align*}
 &\mathrel{\phantom{\Iff}} \mu' \subset \lift{R'}(\uppera{R'}{\mu'}) \\
 &\Oif {\mu'}\comp{i} \subset (\uppera{R'}{\mu'})\comp{R'}\comp{i} \\
 &\Iff \mu \subset \lift{R}(\uppera{R'}{\mu'}) \\
 &\Iff \uppera{R}{\mu} \subset \uppera{R'}{\mu'}
\end{align*}
By the dual discussion, $\lowera{R'}{\mu'} \subset \lowera{R}{\mu}$ can be proved as well.

In addition to those basic results, our theory is useful for guessing decision rules with a non-surjective functor $R$.
For estimation of unknown attribute values, we follow the policy: an element in the lower approximation is not refutable and an element of the upper approximation should have an evidence.
The following example automatically generates decision rules for unknown attribute values.
In this sense, we can say that calculations of our approximations reveal hidden decision rules.

\begin{example}\label{Ex:Preorder}
 Given the same data as Example~\ref{Ex:Simple}.
 Also assume that we know the value ``high'' is more influential for the positive answer than ``low''.\footnote{If we do not know whether such a condition really holds, we can automatically force the approximations to follow it.  Hence, in fact, it is enough for us to believe that the condition holds.}
 We can consider a preordered set $A' = \cset{\text{red},\text{blue},\text{yellow},\text{white}}\times\cset{\text{low},\text{middle},\text{high}}$ instead of $A$, where $\text{low} < \text{middle} < \text{high}$ hold.
 Let $R'$ be a functor from $X$ to $A'$ defined from the data table.
 We can guess the values for ``middle'', which is an intermediate level between ``high'' and ``low''.
 \begin{align*}
  \uppera{R'}{\mu} &= \uppera{R}{\mu} \cup
  \cset{(\text{red},\text{middle}),(\text{yellow},\text{middle})} \\
  \lowera{R'}{\mu} &= \lowera{R}{\mu} \cup
  \cset{(\text{red},\text{middle}),(\text{yellow},\text{middle})}
 \end{align*}
 These approximations do not contradict our intuition.
 For these results, we do not need any special consideration.
 We have just calculated the approximations.

 One may consider the following pair is a preferable solution for ``middle''.
 \begin{align*}
  &\uppera{R}{\mu} \cup
  \cset{(\text{red},\text{middle}),(\text{yellow},\text{middle}),(\text{white},\text{middle})} \\
  &\lowera{R}{\mu} \cup
  \cset{(\text{red},\text{middle})}
 \end{align*}
 In this approach, an element of the lower approximation requires non-empty positive support, and an element of the upper approximation should not be prohibited.
 On the other hand, in our approach, all non-refutable elements are in the lower approximation, and an element of the upper approximation requires an explicit evidence.
\end{example}

Preorders can be used also in analysis of incomplete information systems~\cite{Kryszkiewicz98:RSAIIS}.
If a top element $\top$ exists in $A$, we have the following properties.
\begin{align*}
 \top \in{\uppera{R}{\mu}} &\Iff \myforsome{x \in{X}} Rx \in{\uppera{R}{\mu}}  \\
 \top \in{\lowera{R}{\mu}} &\If  \myforsome{x \in{X}} Rx \in{\lowera{R}{\mu}}
\end{align*}
This fact suggests that a `missing' value can be represented as $\top$.
The first statement is not so much meaningful because $\top$ is always included in $\uppera{R}{\mu}$ for a non-empty $\mu$.
Our theory gives decision rules for incomplete data different from those of the previous studies~\cite{Kryszkiewicz98:RSAIIS,StefanowskiTsoukias01:IITRC}.
The dual notion of a missing value is a `wild card'.
A wild card can be instantiated into any value.
For a bottom element $\bot$ in $A$, the following statements hold.
Note that $\bot \in{\lowera{R}{\mu}}$ is equivalent to $\mu = X$.
\begin{align*}
 \bot \in{\uppera{R}{\mu}} &\Oif \myforall{x \in{X}} Rx \in{\uppera{R}{\mu}}  \\
 \bot \in{\lowera{R}{\mu}} &\Iff \myforall{x \in{X}} Rx \in{\lowera{R}{\mu}}
\end{align*}
In this sense, a wild card can be interpreted as $\bot$ in $A$.
In our framework, data including $\top$ affect lower approximations and data including $\bot$ affect upper approximations.
For practical applications, a top/bottom element in a subspace of $A$ rather than $A$ itself is useful.
There is room for discussion in effectiveness of this method.

\section{Refinement of Rough Sets}
\label{Sec:Refined}

In this section, we replace $2$ with a general complete lattice $L$.
Let $L^{X}$ be the set of all monotone functions from $X$ to $L$.
It can be seen easily that $L^{X}$ is a preordered set, where $\mu \leq \mu'$ is defined as $\mu{x} \leq \mu'{x}$ for any $x \in{X}$.
In $2^{X}$, ${\leq}$ is just the same as ${\subset}$.
For any functor $R \dom{X}{A}$, $\lift{R} \dom{L^{A}}{L^{X}}$ can be defined in the same way as the case of $2$.

\begin{definition}
 For a functor $R \dom{X}{A}$ between preordered sets, the \emph{refined upper approximation} $\uppera{R} \dom{L^{X}}{L^{A}}$ is a left adjoint of $R^{\ast}$, and the \emph{refined lower approximation} $\lowera{R} \dom{L^{X}}{L^{A}}$ is a right adjoint of $R^{\ast}$.
\end{definition}

Without any difficulty, we have the same results for existence of refined approximations as in the previous section.
The completeness condition of $L$ is required for existence of right-hand sides of the equations.

\begin{theorem}\label{Thm:ExExt}
 The refined upper/lower approximation exists uniquely for any $R \dom{X}{A}$.
 The pointwise definitions of the approximations are as follows.
 \begin{align*}
  (\uppera{R}{\mu}){a} &= \join{\sset{\mu{x}}{R{x} \leq a}} \\
  (\lowera{R}{\mu}){a} &= \meet{\sset{\mu{x}}{a \leq R{x}}}
 \end{align*}
\end{theorem}

Also reducibility and updates are defined in the same way.
All the results in Section~\ref{Sec:Simple} hold for refined approximations, replacing $\subset$ with $\leq$.

\begin{remark}
 In the case $L$ is $\interval{0}{1}$, a pair of refined upper and lower approximations are sometimes called a rough fuzzy set~\cite{DuboisPrade90:RFSFRS}.
 As the notion of rough fuzzy sets is different from the notion of fuzzy rough sets, enriched approximations in the next section is more general than refined approximations in this section.
\end{remark}

Usually, when the cardinality of the decision attribute is more than two, we reduce it to $2$ before calculating approximations.
Such a process can be justified in the rest of this section.

Before changing answer sets, we transform the definition of approximations.
It is well-known that a left/right adjoint on function objects induces a left/right Kan extension.
Whenever you need, it is possible to define approximations as Kan extensions.

\begin{lemma}\label{Thm:Kan}
 For any functors $R \dom{X}{A}$ and $\mu \dom{X}{L}$, $\uppera{R}{\mu}$ is a left Kan extension of $\mu$ along $R$.
 Dually, $\lowera{R}{\mu}$ is a right Kan extension.
\end{lemma}

Diagram representations may be helpful to understand the situation.
$\uppera{R}{\mu}$ is the minimum in the following $U$'s and $\lowera{R}{\mu}$ is the maximum in $D$'s.
\begin{gather*}
 \xymatrix{
  {X} \ar@{{}{}{}}[rd]|(.3){\leq} \ar[r]^-{R} \ar[d]_-{\mu} &
  {A} \ar[ld]^-{U}
  & {} & {} &
  {X} \ar@{{}{}{}}[rd]|(.3){\geq} \ar[r]^-{R} \ar[d]_-{\mu} &
  {A} \ar[ld]^-{D}
  \\
  {L} & {}
  & {} & {} &
  {L} & {}
 }
\end{gather*}

The following lemma is known as a general property of Kan extensions.

\begin{lemma}\label{Thm:KanPrsrv}
 If a functor $H \dom{L}{L'}$ has a left adjoint, it preserves right Kan extensions, that is, ${H}\comp{\lowera{R}{\mu}} = \lowera{R}({H}\comp{\mu})$ holds.
 The dual property holds for left Kan extensions.
\end{lemma}

Consider the functor $H^{q} \dom{L}{2}$ defined below.
\begin{align*}
 H^{q}{p} &= \top \textif q \leq p \\
 &= \bot \otherwise
\end{align*}
A left adjoint of this $H^{q}$ is ${\blank}\wedge{q}$, where $2$ is embedded to $L$ implicitly.
Therefore, $H^{q}$ preserves right Kan extensions.
While $H^{q}$ is practically useful, the dual of $H^{q}$ is a little tricky.
Let $\lnot{H_{q}}$ be defined as follows.
\begin{align*}
 \lnot{H_{q}}{p} &= \bot \textif p \leq q \\
 &= \top \otherwise
\end{align*}
Since ${\blank}\vee{q}$ is a right adjoint of $\lnot{H_{q}}$, $\lnot{H_{q}}$ preserves left Kan extensions.
Although $H^{q}$ does not preserve upper approximations in general, $\uppera{R}({H^{q}}\comp{\mu}) \leq {H^{q}}\comp{\uppera{R}{\mu}}$ still holds by the definition of $\uppera{R}$.

If the decision attribute is discrete, the situation is simpler.
Let $X$ and $D$ be discrete sets and $\mu$ be a usual function from $X$ to $D$.
We can define a complete lattice by adding $\bot$ and $\top$ to $D$.
In this case, $q \leq p$ is equivalent to $q = p$ for $p,q \in{S}$.
Since $H^{q}$ preserves lower approximations, $q = (\lowera{R}{\mu})(R{x})$ means $q = \mu{y}$ for any $y$ in the equivalence class $R{x}$.
On the other hand, $\lnot{H_{q}}$ intuitively means the predicate, an input is not equal to $q$.
So, the preservation by $\lnot{H_{q}}$ induces the fact that an upper approximation is essentially a lower approximation in the discrete case.
Indeed, $(\uppera{R}{\mu})(R{x}) = (\lowera{R}{\mu})(R{x})$ holds if $(\uppera{R}{\mu})(R{x})$ is not $\top$ nor $\bot$.
If $(\uppera{R}{\mu})(R{x})$ is $\top$ or $\bot$, $(\lowera{R}{\mu})(R{x})$ is its negation.

Such undesirable situation is caused by the poor structure of $\cset{\bot,\top}\cup{S}$.
Instead, we can consider a lattice $L$ freely generated from $D$.
If we assume $D$ is finite, $L$ is complete.
Though $\lnot{H_{q}}$ is still useless for $L$, we can prove algebraically that $H^{q}$ preserves upper approximations.
This preservation follows from the property of free lattices: if we take $p$ and $p_i$ in $D$, $p \leq \join{p_i}$ implies $p = p_i$ for some $i$.

\begin{theorem}
 Let $L$ be a free lattice over a finite discrete set $D$.
 For any $q \in{S}$, not only ${H^{q}}\comp{\lowera{R}{\mu}} = \lowera{R}({H^{q}}\comp{\mu})$ but also ${H^{q}}\comp{\uppera{R}{\mu}} = \uppera{R}({H^{q}}\comp{\mu})$ hold.
\end{theorem}

\section{Enriched Rough Sets}
\label{Sec:Enriched}

In the previous section, $2$ in the codomain has been replaced with a complete lattice.
In this section, deeper replacement of $2$ is discussed.
It is well-known that a preordered set is a $2$-enriched category and a monotone function between preordered sets is a $2$-enriched functor.
Therefore, we can generalize the theory of rough sets in the manner of enriched category theory.
We emphasize again that all the theorems in this section can be derived from known general theorems without any cost.

\begin{definition}
 A lattice $L$ with an extra monoid structure $\tuple{1,{\copower}}$ is called a \emph{residuated lattice} if every ${p}\copower{\blank}$ and ${\blank}\copower{p}$ have right adjoints.
 If ${\copower}$ is commutative, we say that $L$ is commutative.
 If $L$ is complete as a lattice, we say that $L$ is complete.
 We assume that all residuated lattices in this paper are commutative and complete.
\end{definition}

Though the definition of residuated lattices does not require the monotonicity of ${\copower}$, it follows from properties of adjunctions.
It is also provable that ${\copower}$ is distributive over ${\vee}$.

\begin{remark}
 We have chosen commutative residuated lattices for enrichment because the standard theory~\cite{Kelly82:BCECT} assumes a symmetric monoidal closed structure.
 The use of a residuated lattice for a generalization of fuzzy rough sets~\cite{RadzikowskaKerre04:FRSBRL,SheWang09:AAFRSBRL} is not a novel idea, though it is a category theoretic requirement in our framework.
 In a fuzzy setting, more relaxed structures may be preferred.
 Though we do not aim at such generalization, some part of this section can be enriched by a quantaloid~\cite{Stubbe05:CSEQ,Stubbe06:CSEQ}.
\end{remark}

\begin{remark}
 A residuated lattice is also known as an algebraic model of a kind of substructural logic, though we do not mention the details in this paper.
\end{remark}

\begin{example}
 Since a Heyting algebra is a commutative residuated lattice, we can take a complete Heyting algebra as $L$.
 Especially, $\interval{0}{1}$ is a complete commutative residuated lattice with the usual order and the minimum operator.
 In this case, $r^{p}$ is $r$ if $p \leq r$, otherwise $\top$.
\end{example}

\begin{example}
 We can consider a non-canonical monoid structure in $\interval{0}{1}$.
 It is obvious that $\interval{0}{\infty}$ is a complete commutative residuated lattice with the canonical order and multiplication.
 Since $\interval{0}{\infty}$ is isomorphic to $\interval{0}{1}$ as ordered sets, we can introduce a monoid structure into $\interval{0}{1}$.
\end{example}

\begin{definition}
 Let $L$ be a complete commutative residuated lattice with an order ${\preceq}$ and a monoid structure $\tuple{1,{\copower}}$.
 Let $\exponent{p}{\blank}$ be the right adjoint of ${p}\copower{\blank}$, that is, ${p}\copower{q} \preceq r$ is equivalent to $q \preceq \exponent{p}{r}$.
 A (small) \emph{$L$-enriched category} is a set $X$ with an $L$-indexed relation ${\leq}$ on $X$ satisfying the following conditions.
 \begin{itemize}
  \item For any $x$ and $y$, there is a unique value $p \in{L}$ such that $x \eleq{p} y$.
  \item $x \eleq{p} x$ implies $1 \preceq p$.
  \item $x \eleq{p} y$, $y \eleq{q} z$ and $x \eleq{r} z$ imply ${q}\copower{p} \preceq r$.
 \end{itemize}
\end{definition}

\begin{remark}
 One may prefer a notation like $p = {\mathrm{Hom}}(x,y)$ to the above notation $x \eleq{p} y$.
 It is more familiar to category theory, but we put a priority on compatibility with the previous sections.
\end{remark}

\begin{remark}
 In any $L$-enriched category, $\union\sset{{\eleq{p}}}{1 \preceq p}$ is a preorder.
 In this sense, an $L$-enriched category is a preordered set.
 Meets and joins can be defined in $X$ with respect to this preorder.
 In the case that $L$ is $2$, $x \eleq{\bot} y$ is often written as $x \not\leq y$.
\end{remark}

\begin{remark}
 For a preordered set $X$, we can define enrichment on $X$: $x \eleq{1} y$ if $x \leq y$, otherwise $x \eleq{\bot} y$.
 In this way, also a discrete set can be regarded as an $L$-enriched category.
\end{remark}

\begin{definition}
 For $L$-enriched categories $X$ and $Y$, an \emph{$L$-enriched functor} $F \dom{X}{Y}$ is a function on objects satisfying the following condition.
 \begin{itemize}
  \item $x \eleq{p} x'$ and $F{x} \eleq{q} F{x'}$ imply $p \preceq q$.
 \end{itemize}
 The set of $L$-enriched functors from $X$ to $Y$ is denoted by $Y^{X}$.
 For $L$-enriched functors $F,G \dom{X}{Y}$, $F \nattr G$ means that $F{x} \eleq{p} G{x}$ implies $1 \preceq p$.
 When $F \nattr G$ holds, we may say that there is an \emph{$L$-enriched natural transformation} from $F$ to $G$.
\end{definition}

\begin{lemma}\label{Thm:DefEnr}
 If functions on objects $F \dom{X}{Y}$ and $H \dom{Y}{Z}$ are $L$-enriched functors, ${H}\comp{F}$ is also an $L$-enriched functor.
 For $L$-enriched functors $F,G \dom{X}{Y}$, $H \dom{Y}{Z}$ and $H' \dom{W}{X}$, $F \nattr G$ implies ${H}\comp{F} \nattr {H}\comp{G}$ and ${F}\comp{H'} \nattr {G}\comp{H'}$.
\end{lemma}

The category whose object is an $L$-enriched category and whose morphism is an $L$-enriched functor is denoted by $\cats{L}$.
In this paper, $\cats{L}$ itself is not an enriched category but just a category in the usual sense.

\begin{lemma}\label{Thm:BaseEnr}
 $L$ can be an $L$-enriched category with the enrichment $p \eleq{\exponent{p}{q}} q$.
\end{lemma}

\begin{lemma}\label{Thm:FunEnr}
 For $L$-enriched categories $X$ and $Y$, $Y^{X}$ can be $L$-enriched so that $F \eleq{q} G$ implies $q = \meet{\sset{p_x}{F{x} \eleq{p_x} G{x}}}$.
 Especially, $L^{X}$ is an $L$-enriched category.
\end{lemma}

\begin{definition}
 For $L$-enriched categories $X$ and $Y$, an $L$-enriched category ${X}\otimes{Y}$ is defined as follows.
 \begin{itemize}
  \item An object of ${X}\otimes{Y}$ is a pair of objects of $X$ and $Y$.
  \item $(x,y) \eleq{r} (x',y')$ in ${X}\otimes{Y}$ implies $r = {p}\copower{q}$ for some $p$ and $q$ such that $x \eleq{p} x'$ and $y \eleq{q} y'$ hold.
 \end{itemize}
 Similarly, ${X}\times{Y}$ is defined.
 \begin{itemize}
  \item An object of ${X}\times{Y}$ is a pair of objects of $X$ and $Y$.
  \item $(x,y) \eleq{r} (x',y')$ in ${X}\times{Y}$ implies $r = {p}\wedge{q}$ for some $p$ and $q$ such that $x \eleq{p} x'$ and $y \eleq{q} y'$ hold.
 \end{itemize}
\end{definition}

\begin{lemma}\label{Thm:TensorEnr}
 In $\cats{L}$, ${X}\otimes{\blank}$ is a left adjoint of ${\blank}^{X}$, and ${X}\times{Y}$ is a product of $X$ and $Y$.
\end{lemma}

\begin{remark}
 In general, ${X}\otimes{Y}$ is not a product of $X$ and $Y$.
 If $L$ is a complete Heyting algebra, that is, ${\copower}$ coincides with ${\wedge}$, $\cats{L}$ is cartesian closed.
\end{remark}

For an $L$-enriched functor $R \dom{X}{A}$, the $L$-enriched functor $\lift{R} \dom{L^{A}}{L^{X}}$ is defined in the same way as on preordered sets.
We have to check $\lift{R}$ is really an $L$-enriched functor.
Suppose $\mu \eleq{p} \nu$ in $L^{A}$.
$p$ is $\meet{\sset{p_a}{\mu{a} \eleq{p_a} \nu{a}}}$ by definition.
Since $\lift{R}{\mu} \eleq{q} \lift{R}{\nu}$ implies $q = \meet{\sset{q_x}{\mu(R{x}) \eleq{q_x} \nu(R{x})}}$, we can conclude $p \preceq q$.

\begin{definition}
 Given $L$-enriched functors $F \dom{X}{Y}$ and $G \dom{Y}{X}$.
 $F$ is a \emph{left ($L$-enriched) adjoint} of $G$ and $G$ is a \emph{right ($L$-enriched) adjoint} of $F$ if the following condition is satisfied.
 \begin{itemize}
  \item For any $x \in{X}$ and $y \in{Y}$, $F{x} \eleq{p} y$ is equivalent to $x \eleq{p} G{y}$.
 \end{itemize}
\end{definition}

\begin{definition}
 For an $L$-enriched functor $R \dom{X}{A}$, the \emph{$L$-enriched upper approximation} $\uppera{R} \dom{L^{X}}{L^{A}}$ is a left $L$-enriched adjoint of $R^{\ast}$, and the \emph{$L$-enriched lower approximation} $\lowera{R} \dom{L^{X}}{L^{A}}$ is a right $L$-enriched adjoint of $R^{\ast}$.
\end{definition}

Due to the general theory of enriched categories, we can get existence of approximations.
Pointwise approximations can be calculated via enriched Kan extensions.
Like the preordered case, an upper approximation is a left Kan extension, and a lower approximation is a right Kan extension in the sense of $L$-enriched theory.

\begin{theorem}\label{Thm:ExEnr}
 The $L$-enriched upper/lower approximation exists uniquely for any $R \dom{X}{A}$.
 The pointwise definitions of the approximations are as follows.
 \begin{align*}
  (\uppera{R}{\mu}){a} &= \join{\sset{{p}\copower{\mu{x}}}{R{x} \eleq{p} a}} \\
  (\lowera{R}{\mu}){a} &= \meet{\sset{\exponent{p}{(\mu{x})}}{a \eleq{p} R{x}}}
 \end{align*}
\end{theorem}

The notions of reduction and update can be defined straightforwardly for $L$-enriched categories.

\begin{definition}
 Given $L$-enriched functors $R \dom{X}{A}$ and $\mu \dom{X}{L}$.
 $A$ is \emph{reducible} to $A'$ along $R' \dom{A}{A'}$ if $\lift{R'}(\uppera{{R'}\comp{R}}{\mu}) = \uppera{R}{\mu}$ and $\lift{R'}(\lowera{{R'}\comp{R}}{\mu}) = \lowera{R}{\mu}$ hold.
\end{definition}

\begin{definition}
 A morphism in the comma category ${\cats{L}}/({A}\times{L})$ is called an \emph{update}.
\end{definition}

$L$-enriched approximations are trivially compositional by definition.
The update property for $L$-enriched approximations also holds.

\begin{theorem}\label{Thm:CompEnr}
 For any $L$-enriched functors $R \dom{X}{A}$ and $R' \dom{A}{A'}$, ${\uppera{R'}}\comp{\uppera{R}} = \uppera{{R'}\comp{R}}$ and ${\lowera{R'}}\comp{\lowera{R}} = \lowera{{R'}\comp{R}}$ hold.
\end{theorem}

\begin{theorem}\label{Thm:RedCompEnr}
 Given $L$-enriched functors $R \dom{X}{A}$ and $\mu \dom{X}{L}$.
 If $A$ is reducible to $A'$ along $R'$ and $A'$ is reducible to $A''$ along $R''$, $A$ is reducible to $A''$ along ${R''}\comp{R'}$.
\end{theorem}

\begin{theorem}\label{Thm:RefineEnr}
 For any $L$-enriched update $i \dom{X}{X'}$ as a morphism from $\tuple{R,\mu}$ to $\tuple{R',\mu'}$, $\uppera{R}{\mu} \nattr \uppera{R'}{\mu'}$ and $\lowera{R'}{\mu'} \nattr \lowera{R}{\mu}$ hold.
\end{theorem}

We can recall that an $L$-enriched category is preordered and the notions of meets and joins are well-defined in it.
In fact, the preorder in $L^{X}$ is just ${\nattr}$.
In enriched category theory, ${x}\wedge{y}$ is a kind of limit and ${x}\vee{y}$ is a kind of colimit.
It is known that a right $L$-enriched adjoint preserves limits.
So, any lower approximation functor is distributive over ${\wedge}$.
Dually, an upper approximation functor preserves ${\vee}$.

\begin{theorem}\label{Thm:PrsrvEnr}
 Let $\mu$ and $\nu$ be objects of $L^{X}$.
 ${\mu}\wedge{\nu}$ and ${\mu}\vee{\nu}$ exist in $L^{X}$, and moreover, $\lowera{R}({\mu}\wedge{\nu}) = (\lowera{R}{\mu})\wedge(\lowera{R}{\nu})$ and $\uppera{R}({\mu}\vee{\nu}) = (\uppera{R}{\mu})\vee(\uppera{R}{\nu})$ hold.
\end{theorem}

As mentioned before, the $2$-enriched case provides the theory for preordered sets in Section~\ref{Sec:Simple}.
Moreover, the $\interval{0}{1}$-enriched case can provide a theory of fuzzy rough sets.
We show a connection between our theory and other fuzzy theories.
Some studies on fuzzy rough sets and fuzzy Galois connections~\cite{Belohlavek99:FGC,YaoLu09:FGCFP,SheWang09:AAFRSBRL} require that $1$ and $\top$ coincide, but our theory does not assume such a condition.

We can recall that the image of $R \dom{X}{A}$ is isomorphic to $X/{\equiv}$ in the discrete setting.
This fact suggests that surjective $R$ can be constructed from a binary relation ${\equiv}$.
\begin{align*}
 &\mathrel{\phantom{\Iff}} {\equiv} \dom{{X}\times{X}}{2} \\
 &\Iff \curry{\equiv} \dom{X}{2^{X}} \\
 &\Oif \curry{\equiv} \dom{X}{X/{\equiv}}
\end{align*}
Here, $\curry{f}$ is the curryed form of $f$, that is, $(\curry{f}{y}){x} = f(x,y)$ holds.
The second inference is just restricting codomain to the image.
Though $\curry{\equiv}$ sends $x$ to its equivalence class, it is not essential for the construction that ${\equiv}$ is an equivalence relation.
We generalize the above construction to $L$-enriched categories.
Given an $L$-enriched functor ${\equiv} \dom{{X}\otimes{X}}{L}$, which is referred to as an \emph{$L$-enriched binary relation}.
We can construct $R = \curry{\equiv}$ as follows.
\begin{align*}
 &\mathrel{\phantom{\Iff}} {\equiv} \dom{{X}\otimes{X}}{L} \\
 &\Iff \curry{\equiv} \dom{X}{L^{X}} \\
 &\Oif \curry{\equiv} \dom{X}{X/{\equiv}}
\end{align*}
The first correspondence can be derived from the adjunction of Lemma~\ref{Thm:TensorEnr}.
We call $\curry{\equiv}{x}$ the $L$-enriched equivalence class of $x$.
${X}/{\equiv}$ is the full subcategory of $L^{X}$, which consists of equivalence classes.
We can note that the inclusion from ${X}/{\equiv}$ to $L^{X}$ is conservative with respect to $L$-enriched approximations.
If we apply our enriched theory to this $R$, we can get the standard theory of fuzzy rough sets.

If the following conditions are satisfied, ${\equiv} \dom{{X}\otimes{X}}{L}$ is called an \emph{$L$-enriched equivalence relation}.
\begin{align*}
 &1 \preceq ({x}\mathbin{\equiv}{x}) \\
 &({x}\mathbin{\equiv}{y})\copower({y}\mathbin{\equiv}{z})
 \preceq ({x}\mathbin{\equiv}{z}) \\
 &({x}\mathbin{\equiv}{y}) = ({y}\mathbin{\equiv}{x})
\end{align*}
When ${\equiv}$ is an equivalence relation, $\curry{\equiv}x \eleq{p} \curry{\equiv}y$ implies $p = ({x}\mathbin{\equiv}{y})$.
Hence, the following equations hold.
\begin{align*}
 (\uppera{\curry{\equiv}}{\mu})(\curry{\equiv}z) &=
 \join_{x\in{X}}{({x}\mathbin{\equiv}{z})\copower{\mu{x}}} \\
 (\lowera{\curry{\equiv}}{\mu})(\curry{\equiv}z) &=
 \meet_{x\in{X}}{\exponent{({z}\mathbin{\equiv}{x})}{(\mu{x})}}
\end{align*}
In this sense, the usual theory of fuzzy rough sets is an instance of our $L$-enriched theory.

Practically, we first fix a binary relation on $\underline{A}$, i.e., a functor $S \dom{{\underline{A}}\otimes{\underline{A}}}{L}$, where $\underline{A}$ is the underlying set of $A$.
If $S$ is an $L$-enriched equivalence relation, trivially $S$ gives an $L$-enrichment of $A$, i.e., ${a}\eleq{p}{b}$ is given as $({a}\mathbin{\equiv}{b}) = p$.
With this enrichment, $S$ can be regarded as an $L$-enriched relation on $A$.
This fact follows the enriched version of Yoneda's lemma.

\begin{definition}
 Let $A$ be an $L$-enriched category.
 $\opp{A}$ denotes the \emph{opposite category} of $A$: $a \eleq{p} b$ in $\opp{A}$ is just $b \eleq{p} a$ in $A$.
 Since $L$ is commutative, $\opp{A}$ is trivially an $L$-enriched category.
 We define the $L$-enriched functors $H^{a} \dom{A}{L}$ and $H_{a} \dom{\opp{A}}{L}$ as follows.
 \begin{align*}
  H^{a}{b} &= p \textif a \eleq{p} b \\
  H_{a}{b} &= p \textif b \eleq{p} a
 \end{align*}
 The \emph{Yoneda embedding} from $A$ to $L^{\opp{A}}$ is the $L$-enriched functor that sends $a$ to $H_{a}$.
\end{definition}

It is well-known that any $L$-enriched category $A$ can be embedded into $L^{\opp{A}}$ via the Yoneda embedding.
The following lemma is known as a corollary of Yoneda's lemma.

\begin{lemma}
 The Yoneda embedding is really an embedding, that is, the following condition holds.
 \begin{itemize}
  \item If $a \eleq{p} a'$ holds, $H_{a} \eleq{p} H_{a'}$ and $H^{a'} \eleq{p} H^{a}$ hold.
 \end{itemize}
\end{lemma}

If $A$ is symmetric, that is, $\opp{A}$ is $A$, the Yoneda embedding canonically provides an $L$-enriched equivalence relation on $A$.
When the enrichment of $A$ is given by $S \dom{{\underline{A}}\otimes{\underline{A}}}{L}$, the Yoneda embedding coincides with $S$ on objects.
Therefore, an $L$-enriched equivalence relation $S$ on $\underline{A}$ canonically determines $L$-enrichment of $A$ and $S$.

In the avobe discussion, the symmetricity condition can be removed though we have to consider the opposite categories.
Such generalization is needed for asymmetric enrichment like Example~\ref{Ex:Preorder}.

If an $L$-enriched equivalence relation $S \dom{{A}\otimes{A}}{L}$ is given, we can define $x \equiv y$ as $S(R{x},R{y})$ for an arbitrary $R \dom{X}{A}$.
This ${\equiv}$ is an $L$-enriched equivalence relation as well.
As expected, ${X}/{\equiv}$ is isomorphic to $A$ when $R$ is surjective on objects and $\curry{S}$ is injective on objects.
The latter condition is not so serious: if we take $A/S$ as $A$, the condtition can be dropped.

\begin{example}
 Let $A$ be a metric space with a distance function $d$.
 A set function $S \dom{{A}\times{A}}{\interval{0}{1}}$ can be defined so that $S(a,b) = 2^{-d(a,b)}$.
 If we take $\interval{0}{1}$ with the canonical order and multiplication for $L$, $A$ can be $L$-enriched.
\end{example}

\begin{example}\label{Ex:Enriched}
 Fix $\interval{0}{1}$ with the canonical order and multiplication as $L$.
 Here, $r^{p}$ is $\min(1,r/p)$.
 We give $A_1$ whose underlying set is $\cset{\text{red},\text{blue},\text{yellow},\text{white}}$ and whose enrichment is derived from $S_1 \dom{{A_1}\otimes{A_1}}{\interval{0}{1}}$ below.
 \begin{center}
  \begin{tabular}{c|C{10mm}|C{10mm}|C{10mm}|C{10mm}}
   attribute 1 & red & blue & yellow & white \\ \hline
   red     & $1$   & $1/9$ & $1/2$ & $1/3$ \\ \hline
   blue    & ---   & $1$   & $1/9$ & $1/3$ \\ \hline
   yellow  & ---   & ---   & $1$   & $1/3$ \\ \hline
   white   & ---   & ---   & ---   & $1$
  \end{tabular}
 \end{center}
 Since $S_1$ is symmetric, the left lower triangle remains blank in the table.
 One can see that $S_1$ is indeed an $L$-enriched equivalence relation.
 $A_2$ is a discrete category $\cset{\text{low},\text{high}}$.
 $A_3$ is defined from the following table.
 \begin{center}
  \begin{tabular}{c|C{10mm}|C{10mm}|C{10mm}}
   attribute 3 & iris & freesia & daisy   \\ \hline
   iris    & $1$   & $1$   & $1/9$ \\ \hline
   freesia & ---   & $1$   & $1/9$ \\ \hline
   daisy   & ---   & ---   & $1$
  \end{tabular}
 \end{center}
 Let $A$ be ${A_1}\times{A_2}$, $A'$ be ${A_1}\times{A_2}\times{A_3}$, and $X$ be $\cset{0,1,\dots,11}$.
 The data are given as follows.
 \begin{center}
  \begin{tabular}[t]{c|c|c|c|c}
   object & attribute 1 & attribute 2 & attribute 3 & certainty \\ \hline \hline
   $0$  & red    & low  & iris    & $90/100$ \\ \hline
   $1$  & red    & low  & freesia & $80/100$ \\ \hline
   $2$  & blue   & low  & daisy   & $10/100$ \\ \hline
   $3$  & yellow & low  & daisy   & $35/100$ \\ \hline
   $4$  & yellow & low  & daisy   & $40/100$ \\ \hline
   $5$  & white  & low  & iris    & $25/100$ \\ \hline
   $6$  & red    & high & freesia & $95/100$ \\ \hline
   $7$  & blue   & high & daisy   & $15/100$ \\ \hline
   $8$  & blue   & high & iris    & $15/100$ \\ \hline
   $9$  & yellow & high & freesia & $65/100$ \\ \hline
   $10$ & white  & high & daisy   & $90/100$ \\ \hline
   $11$ & white  & high & daisy   & $70/100$
  \end{tabular}
 \end{center}
 $R \dom{X}{A}$, $R' \dom{X}{A'}$ and $\mu \dom{X}{L}$ are defined from the data in the usual manner.
 Since the image of $R'$ is reducible to $A$ along the projection, here we only show the $L$-enriched approximations of $\mu$ along $R$.
 \begin{align*}
  (\uppera{R}\mu)(\text{red},\text{low})     &= 90/100 &
  (\uppera{R}\mu)(\text{red},\text{high})    &= 95/100 \\
  (\uppera{R}\mu)(\text{blue},\text{low})    &= 10/100 &
  (\uppera{R}\mu)(\text{blue},\text{high})   &= 30/100 \\
  (\uppera{R}\mu)(\text{yellow},\text{low})  &= 45/100 &
  (\uppera{R}\mu)(\text{yellow},\text{high}) &= 65/100 \\
  (\uppera{R}\mu)(\text{white},\text{low})   &= 30/100 &
  (\uppera{R}\mu)(\text{white},\text{high})  &= 90/100 \\
  (\lowera{R}\mu)(\text{red},\text{low})     &= 70/100 &
  (\lowera{R}\mu)(\text{red},\text{high})    &= 95/100 \\
  (\lowera{R}\mu)(\text{blue},\text{low})    &= 10/100 &
  (\lowera{R}\mu)(\text{blue},\text{high})   &= 15/100 \\
  (\lowera{R}\mu)(\text{yellow},\text{low})  &= 35/100 &
  (\lowera{R}\mu)(\text{yellow},\text{high}) &= 65/100 \\
  (\lowera{R}\mu)(\text{white},\text{low})   &= 25/100 &
  (\lowera{R}\mu)(\text{white},\text{high})  &= 45/100
 \end{align*}
\end{example}

As Section~\ref{Sec:Refined}, we can replace $L^{X}$ with ${L'}^{X}$, where $L'$ is not $L'$-enriched but just $L$-enriched.
In this case, $L'$ needs to have powers and copowers (also called tensors).
Change of base by $H \dom{L'}{L}$ can be described like the following diagram.
\begin{gather*}
 \xymatrix{
  {X} \ar@{{}{}{}}[rd]|(.3){\nattr} \ar[r]^-{R} \ar[d]_-{\mu} &
  {A} \ar[ld]|-{\uppera{}{\mu}} \ar[d]^-{\uppera{}({H}\comp{\mu})}
  & {} & {} &
  {X} \ar@{{}{}{}}[rd]|(.3){\revnattr} \ar[r]^-{R} \ar[d]_-{\mu} &
  {A} \ar[ld]|-{\lowera{}{\mu}} \ar[d]^-{\lowera{}({H}\comp{\mu})}
  \\
  {L'} \ar[r]_-{H} & {L} \ar@{{}{}{}}[lu]|(.3){\revnattr}
  & {} & {} &
  {L'} \ar[r]_-{H} & {L} \ar@{{}{}{}}[lu]|(.3){\nattr}
 }
\end{gather*}
If $H$ has a left (resp.\ right) $L$-enriched adjoint, the lower triangle in the right (resp.\ left) diagram commutes.
Especially, $H^{q'}$ defined above preserves lower approximations since a left adjoint of $H^{q'}$ can be defined with copowers.

We have considered complete commutative residuated lattices for enrichment.
Of course, it is possible to define enriched approximations with more complex categories.
The usual large categories are, however, too general for data-mining.
For example, a $\sets$-enriched category is an ordinary (non-enriched) category, and the theory in this section coincides with just the general theory for pointwise Kan extensions.

\section{Concluding Remarks}

In this paper, we have studied categorical formalization of the theory of rough sets.

First, we have characterized the notion of rough sets in terms of category theory.
In our formulation, approximations are defined as adjunctions between preordered sets.
Also the notions of attribute reduction and data update are defined in terms of category theory.
One of the advantages of our framework is that our characterization is closed under composition.
Due to the closedness, we can classify objects step by step in our theory.
In addition, our theory gives a reasoning for guesses of decision rules for unknown data.
Our theory may be applicable to analysis of incomplete data with wild cards and missing attributes.

Second, we have generalized the above theory in the manner of category theory.
We can easily refine the answer type to a complete lattice.
Such refinement is useful for practical situations.
Since a preordered set is a $2$-enriched category and a monotone functions is a $2$-enriched functor, we can generalize $2$ to a more complex algebra.
This paper used a complete commutative residuated lattice $L$ for enrichment.
We have defined the $L$-enriched version of categorical characterization of rough sets.
It has been shown that our enriched theory satisfies the results similar to the theory for preordered sets.
Our theory is not only a generalization of the usual theory of rough sets but also a generalization of the theory of fuzzy rough sets.

\section*{Acknowledgments}

We are grateful to Yuichi Nishiwaki for fruitful discussions.
We would like to thank an anonymous referee for suggesting us the studies on quantaloid-enriched categories.

\bibliography{refs}

\end{document}